\documentclass[fleqn,10pt]{wlscirep}
\usepackage[utf8]{inputenc}
\usepackage[T1]{fontenc}
\usepackage{lineno}
\usepackage{amsmath}
\usepackage{graphicx}
\usepackage{verbatim}
\usepackage{siunitx}
\usepackage{xr}

\makeatletter
\newcommand*{\addFileDependency}[1]{
  \typeout{(#1)}
  \@addtofilelist{#1}
  \IfFileExists{#1}{}{\typeout{No file #1.}}
}
\makeatother

\newcommand*{\myexternaldocument}[1]{%
    \externaldocument{#1}%
    \addFileDependency{#1.tex}%
    \addFileDependency{#1.aux}%
}

\myexternaldocument{supp}

\title{Scenarios of future Indian electricity demand accounting for space cooling and electric vehicle adoption}

\author[1]{Marc Barbar}
\author[1]{Dharik Mallapragada}
\author[1]{Meia Alsup}
\author[1]{Robert Stoner}
\affil[1]{MIT Energy Initiative, Cambridge, 02139, USA}

\begin{abstract}
India is expected to witness rapid growth in electricity use over the next two decades. Here, we introduce a custom regression model to project electricity consumption in India over the coming decades, which includes a bottom-up estimate of electricity consumption for two major growth drivers, air conditioning, and vehicle electrification. The model projections are available at a customizable level of spatial aggregation at an hourly temporal resolution, which makes them useful as inputs to long-term electricity infrastructure planning studies. The approach is used to develop electricity consumption data sets spanning various technology adoption and growth scenarios up to the year 2050 in five-year increments. The aim of the data is to provide a range of scenarios for India's demand growth given new technology adoption. With long-term hourly demand projections serving as an essential input for electricity infrastructure modeling, this data publication enables further work on energy efficiency, generation, and transmission expansion planning for a fast-growing and increasingly important region from a global climate mitigation perspective.
\end{abstract}

\begin{document}

\flushbottom
\maketitle

\thispagestyle{empty}


\section*{Background \& Summary}

\begin{figure}[!ht]
    \centering
    \includegraphics[width=12cm,height=12cm, keepaspectratio]{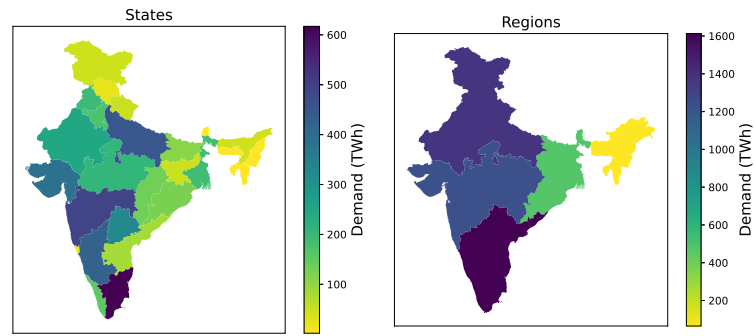}
    \caption{State and regional level distribution of annual electricity 2050 for stable GDP growth, baseline cooling, and home electric vehicle (EV) charging scenario}
    \label{fig:demand}
\end{figure}

\begin{figure}[!ht]
    \centering
    \includegraphics[width=12cm,height=12cm, keepaspectratio]{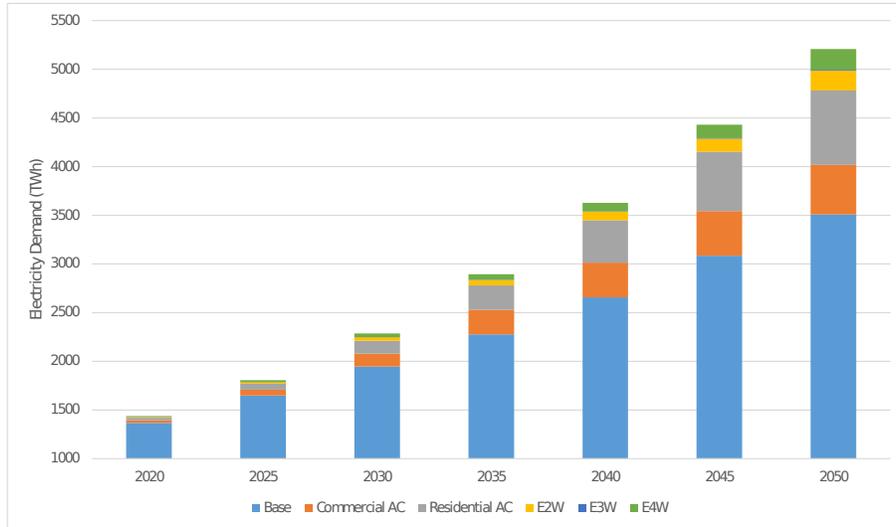}
    \caption{Summary results of India's electricity demand forecasting at national level with stable GDP growth, baseline cooling, and home electric vehicle (EV) charging}
    \label{fig:summary_results}
\end{figure}

\begin{figure}[!ht]
\centering
 \includegraphics[width=7cm,height=7cm, keepaspectratio]{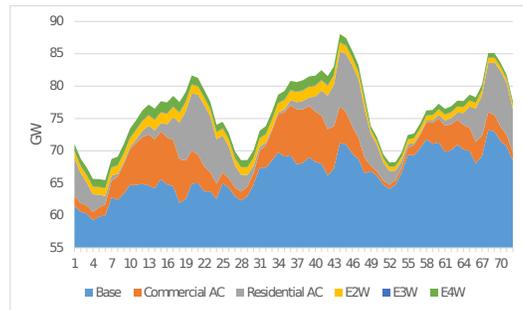}
    \caption{2030 Load profile for southern Region across three days in summer. Scenario: stable GDP growth, baseline cooling, home electric vehicle (EV) charging}
    \label{fig:profile_results}
\end{figure}

Many assessments of future electricity demand in India project large increases in electricity consumption from adoption of air conditioning technologies in the buildings sector over the next two decades \cite{weo20, ev, iea}. This large growth is likely to make India among the top nations in terms of electricity consumption, implying that technology choices related to energy consumption and production in India are likely to play a significant impact on global climate change mitigation efforts. Additionally, the Indian government has been pushing for the transportation sector's electrification, starting with two- and three-wheel vehicles,which is further likely to increase overall electricity demand. As of 2020 in India, there are 152,000 registered electric vehicles \cite{ev}. Air conditioning (AC) related electricity demand accounted for 32.7 TWh, contributing to less than 2.5\% of the total demand in 2019 \cite{iea}. However, both air conditioning and transport electrification are anticipated to introduce structural changes in the temporal and spatial trends in electricity consumption patterns, that has important ramifications for long-term resource planning for the electricity sector \cite{teri}. This paper presents an bottom-up approach to estimate electricity consumption in India for various scenarios of technology and policy adoption with a specific focus on providing aggregated consumption estimates as well as spatio-temporally resolved consumption profiles that would be relevant for regional and national electricity system planning studies. The approach enables quantifying the impact of  various growth and technology adoption scenarios on quantity and pattern in electricity consumption. The datasets detailed in this paper include annual energy consumption at India's state, regional, and national levels as visualized in Fig. \ref{fig:demand}, as well as underlying consumption profiles at an hourly time resolution. The annual energy consumption is forecasted on a five-year increment to 2050. Fig. \ref{fig:summary_results} shows one scenario of national electricity demand forecast. In addition to the snapshot of annual consumption, hourly load profiles are developed at the same resolution as seen in Fig. \ref{fig:profile_results}. 

The forecasting is divided into two steps: business-as-usual and technology. Business-as-usual is a statistical model that infers data it can be trained on i.e. historical electricity demand. The technology model is a bottom-up approach that adds new loads to the total demand. Among new loads, we focus on residential and commercial cooling as well as various electric vehicles (EV). Some key insights from cooling\cite{iea} and EV\cite{ev} studies highlighting peak demand development motivate the need for demand forecasting at the hourly resolution. Cooling demand due to mainly split unit air conditioning installation in India is expected to increase the peak to mean ratio (also sometimes referred to the "peakiness") of electricity demand in India as well as shift the timing of peak demand from evenings to midnight\cite{iea}. While electric vehicles do not constitute a large portion of the total demand, certain charging schemes can contribute significantly to the peak demand\cite{ev}. Numerous energy demand forecast for India have recently been published as decadal snapshots \cite{weo20, teri, brookings}, however granularity of demand at an hourly resolution has not been presented in these studies. Our approach enables quantifying the impact of different technology and structural elements, such as adopting energy efficient vs. baseline cooling technology or work-place charging vs. home charging for EVs, on the hourly electricity consumption profiles. These insights and the accompanying data sets are essential to carry out generation and transmission expansion as well as distribution network planning,and are thus essential for a sustainable energy infrastructure development in the Indian context.

Similar to other forecasting studies, we model Gross domestic product (GDP) growth \cite{mospi} to be the main econometric driver of the business-as-usual demand forecasting, and thus three scenarios are introduced: slow, stable, and rapid GDP growth. We examine two AC load scenarios: energy-efficient equipment and baseline equipment per the International Energy Agency’s Future of Cooling study \cite{iea}. Finally, we evaluate three EV charging mechanisms: home, work, and public charging. This totals the number of data sets spanning three input dimensions to 18 scenarios. Technology adoption growth has been correlated with economic growth under the assumption that new technologies are adopted faster when the economy is growing faster and vice versa. We present two cooling scenarios to highlight the difference in energy-efficient and regular air conditioning units and bring attention to the need for policy and programs that favor energy-efficient cooling unit sales. Furthermore, we present various EV charging mechanisms to inspect the demand impacts that electric vehicle charging can have on the electric grid at different times. The produced data can be used as input to electricity infrastructure planning both at the distribution and transmission level. 

\section*{Methods}
\begin{figure}[!ht]
\centering
 \includegraphics[width=15cm,height=15cm, keepaspectratio]{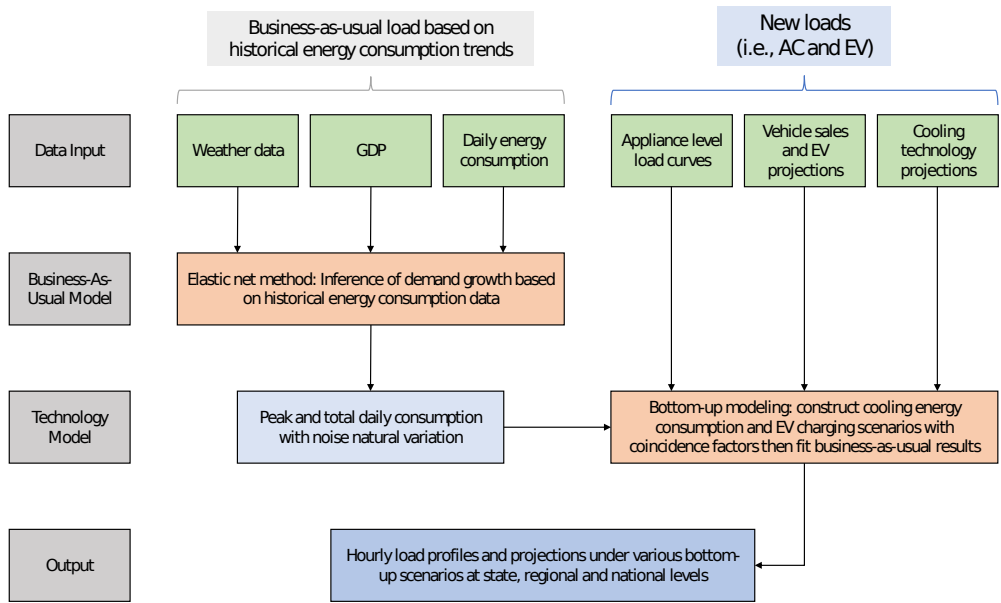}
  \caption{Simplified schematic of methods; inputs in green, models in red, outputs in blue}
   \label{fig:schematic}
\end{figure}

Fig. \ref{fig:schematic} illustrates the major steps of our proposed demand forecasting approach. We use two models to estimate future electricity demand in India. In the first model --- business-as-usual --- we use a linear regression model to project daily peak and consumption on a regional basis; this is the business-as-usual scenario. We then add natural variation to the projections by finding the error between the training data and results and scaling it to every region based on seasonality. Then we fit the projected peak and total consumption to an annual hourly load profile for 2015 \cite{shakti} featuring an evening peak \cite{ivan}. In the second model --- technology model --- we take AC and EV adoption into account as an additive component on top of the business-as-usual predictions. GDP data, which is an independent variable in the model, is chosen to be the main driver of growth of the business-as-usual scenario as well as technology adoption rates. The input data used are publicly available and are referenced in Table \ref{table:data}.

\subsection*{Input data processing}

Although GDP is widely used for forecasting energy demand, it is specifically essential in the case of India, where economic growth is expected to ramp up over the next few decades similar to the recent trends in China \cite{mckinsey}. We based our demand forecast on GDP projections from a PricewaterhouseCoopers (PwC) report \cite{pwc}, that projected India's GDP to grow from 3.6 trillion in 2020 to reach 28 trillion USD in 2050. Considering the historical national GDP data for India starting in 1990, we fit and project an exponential curve for rapid growth and an Gompertz curve for slow growth \cite{gompertz} as detailed in Table \ref{table:gdp}. We use PwC's projections to define the stable GDP growth scenario. Curve fitting and projection results are illustrated in Supplementary Fig. \ref{fig:sup-gdp}. The rapid growth scenario produces an annual average growth rate of 9.5\% , PwC's growth rates start at 7.8\% for the first projected decade and ends at 6.2 \% in the final projected decade. The slow growth scenario starts at 7.2\% growth rate in the first projected decade and ends at 3.9\% in the final projected decade. To break down the regional energy consumption projections to state level we use the ratio of GDP per capita of the corresponding state to the GDP per capita of the region it is in. For each GDP growth scenario, we fit the same functions given state-wise data to produce GDP forecast at the same resolution.  GDP per capita at state-level is computed using the projected GDP data and state level population projections \cite{ssrn}.

\subsubsection*{GDP dependence and limitation}

Relating growth in electricity demand to GDP is a strong generalization, however it is not a novel one in the case of India. Strong correlation between economic growth and energy consumption has been established in the Indian context in this study and other studies \cite{eia} given data from the past two decades \cite{mospi}. We recognize that GDP as a metric of economic growth has several limitations particularly related to projecting how economic growth is distributed among society within a state or nation. This may be the strongest limitation of the data we are presenting in the manuscript. However, lack of historical record and long-term projections of alternative open-access economic data at the desired spatial and temporal resolution limit the development of a framework to project energy consumption with other metrics. While GDP and energy consumption growths may differ in the long-run, there is an evident correlation between the two that can be used to estimate long-run energy consumption growth. Deviating away from linear regression may yield better results, however, data scarcity is again a limitation to the development of more complex models. Furthermore, this manuscript motivates the need for more bottom-up projections and not just regression models because historical consumption cannot infer consumption trends from new demand sources such as cooling and EVs.

Additionally, since the Future of Cooling study by the International Energy Agency relies on GDP forecasts developed by the International Monetary Fund\cite{iea}, we elected to use a similar metric. We intentionally develop a large bandwidth of projection scenarios to mitigate the limitation of an individual snapshot representing a singular assumption. The motivation behind presenting the described results is ability to compare different scenarios and post-analyze the demand growth and the trade-offs. To produce a large bandwidth of growth scenarios we needed to use a straightforward metric that has enough historical data to produce various fitted curves for projections.

\subsection*{Business-as-usual model}

\begin{figure}[!ht]
\centering
 \includegraphics[width=9cm,height=9cm, keepaspectratio]{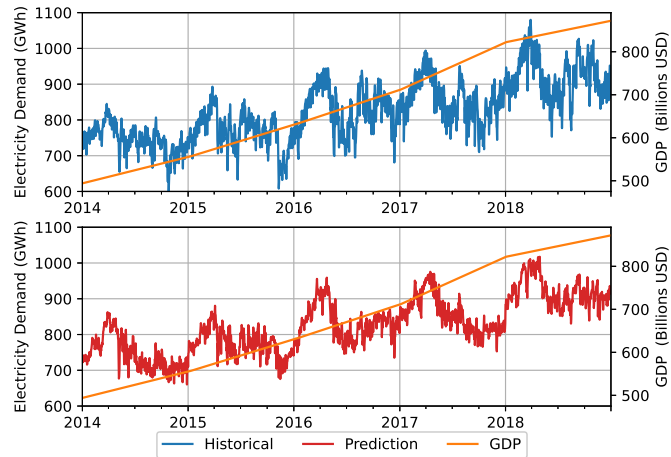}
\caption{Southern region back test annual demand growth given GDP projection}
\label{fig:longrun}
\end{figure}

\begin{figure}[!ht]
\centering
 \includegraphics[width=9cm,height=9cm, keepaspectratio]{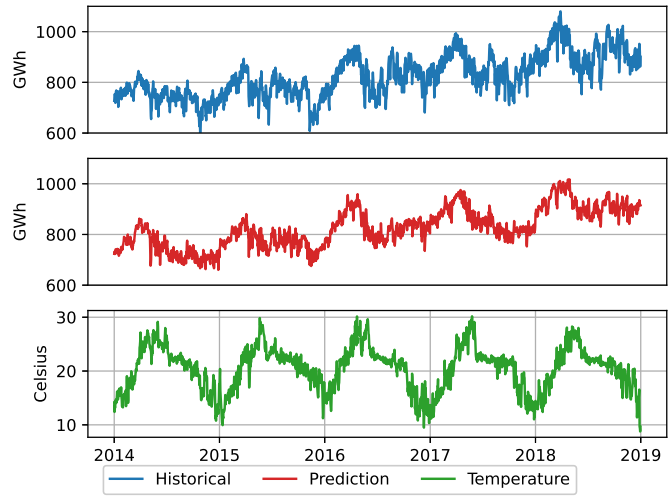}
\caption{Southern region back test seasonal demand variation given weather data}
\label{fig:shortrun}
\end{figure}

The business as usual projections are modeled with a linear regression considering weather and economic growth features. The ground truth historical daily peak and total consumption for each electric grid were obtained from the Power System Operation Corporation (POSOCO) for 2014-2019 \cite{posoco}. The GDP used in the model was obtained, as explained in the previous section. Weather data was secured from the NASA Merra-2 data set \cite{nasa}. The choice of features for the regression model is limited to GDP and weather variation due to the limitation in availability of data, both historical and future projections, at the desired spatial and temporal resolution. GDP is identified as a long-term parameter driving growth in year over year demand projections as highlighted in Fig. \ref{fig:longrun}. Weather data is identified as a short-term parameter driving seasonal variation within a year's demand projections as highlighted in Fig. \ref{fig:shortrun}. Previous parametric analysis on these features and their coefficient for short and long term demand forecasting in both time and frequency domain \cite{meia} reinforce their use as features for the business-as-usual regression model. We present detailed outcomes for the Southern region, with further details available in \cite{meia}.

\subsubsection*{NASA Merra 2 data acquisition}

For each of the five electric grid demand regions highlighted in right panel of Fig. \ref{fig:demand}, the largest cities in each region were identified using population data made available by the United Nations\cite{pop}. Then, the city's latitude and longitude were used to pull down the corresponding environmental data from the Nasa Merra-2 data set. The cities used for each of the five regions are listed here:
\begin{itemize}
    \item Northern: Delhi, Jaipur, Lucknow, Kanpur, Ghaziabad, Ludhiana, Agra
    \item Western: Mumbai, Ahmadabad, Surat, Pune, Nagpur, Thane, Bhopal, Indore, Pimpri-Chinchwad
    \item Eastern: Kolkata, Patna, Ranchi (Howrah was ignored because the environmental factors are the same as Kolkata)
    \item Southern: Hyderabad, Bangalore, Chennai, Visakhapatnam, Coimbatore, Vijayawada, Madurai
    \item Northeast: Guwahati, Agartala, Imphal
\end{itemize}

From the NASA set, 11 variables were included for each city: specific humidity, temperature, eastward wind, and northward wind (all 2m above the surface and 10m above the surface - eight total variables), precipitable ice water, precipitable liquid water, and precipitable water vapor. In particular, the instantaneous two-dimensional collection "inst1\_2d\_asm\_Nx (M2I1NXASM)" from NASA was used. Detailed descriptions of these variables are available in the Merra-2 file specification provided by NASA\cite{nasa} . The environmental variables available from the NASA MERRA-2 dataset were given on an hourly basis.  The daily minimum, daily, maximum, and daily average was calculated for each of the 11 variables for each day.

\subsubsection*{Forecasts}
The business-as-usual demand forecasting problem was divided into ten separate problems,corresponding to one problem each peak and total consumption for each of the five regional grids shown in Figure \ref{fig:demand}.  To ensure the model would not overfit the data, the model was trained with Elastic Net \cite{scikit-learn} to regularize results, and validated on held out 2019 data. An L1 ratio (Lasso) of .9 was chosen to minimize error in 2019 as the validation set. Then all of the models were trained with .9 L1 ratio on the full dataset.

\subsubsection*{Addition of natural variation}
This step aimed to match the statistical characteristics of an actual load year with the projected year. 2019 was used to derive the differences. Natural variation was estimated by a distribution characterized by the mean and standard deviation of the differences (in absolute value). Then, a natural variation adjustment was added to that day (with a random true/false bit for positive or negative variation). The noise was calculated for each region and peak demand and daily consumption separately. The natural variation (noise) vectors used are on the Github repository for this paper \cite{git}. This part of the process is non-deterministic and replication of the results requires using the same natural variation vector used in our projections.

\subsubsection*{Hourly profiles}
The statistical inference model presented above forecasts daily consumption driven by state-level economic parameters and weather data. The produced projections are at a daily resolution. We downscaled the data to hourly load profiles based on the 2015 hourly load profile data \cite{shakti}. The result of the regression model is at regional level, breaking it down state-wise is pro-rated based on state-wise to region-wise GDP per capita projections ratios for the respective year. To do so, we tag each day of the year by the month it corresponds to and whether it is a weekday or weekend. We cluster demand for each hour by month and day. Each hour of the day then has its own cluster of demand data from 2015 based on the assumption that the same hour of the day for a given month and the same day type will exhibit similar demand behavior. This biases the construction of the profiles to demand patterns from 2015 only. To minimize the impact of this bias, we use the historical weather data\cite{nasa} of the testing data years (2014-2019) for each day to simulate daily temperatures variations that are reflected in higher or lower demand. We sample weather data for each day and compare it to 2015, and subsequently use normalized the difference to scale the demand on a daily basis. Finally, we sample demand for each hour of the year from the corresponding cluster (defined by month and weekend or weekday) and scale it accordingly. Constructing the hourly load profile and fitting them to match the projected daily consumption and the projected daily peak demand then becomes a trivial exercise of sampling and fitting from the corresponding clusters and weather data space. The 2015 hourly demand data used in this study is documented in detail elsewhere and has been used in projecting demand for supply-side modeling efforts \cite{ivan}. Limited availability of complete hourly data at state and regional level in India biases the hourly profiles to the 2015 datasets. However, the business-as-usual projections are for existing demands composed mainly of lighting and appliance at the residential level and large daytime loads at the commercial level \cite{usaid}. Our approach implicitly assumes that energy consumption trends for these loads will follow historical patterns and therefore sampling from a given year with post-processed noise variation can yield reasonable results.

\subsubsection*{Impact of Climate change on business-as-usual demand}
As per the International Energy Agency (IEA) World Energy Outlook (WEO) 2019\cite{weo19} only 5\% of households in India currently own air conditioning units and 2.6\% of commercial building energy use is from space cooling. Historically, electricity consumption in India has been driven by lighting and appliances in the residential sector \cite{usaid} with commercial and industrial sector contributing via larger daytime loads. Since cooling demand is not historically available in the data that the business-as-usual regression model is learning from, there is no parametric value to projecting increase in temperatures since there is no evident correlation between temperature increase and lighting or appliance use. Moreover, since space cooling is a small percentage of current electricity demand in India, no major trends can be identified given the limited daily training data that is being used for the business-as-usual regression. It is then safe to assume that weather remains constant for the business-as-usual demand.

\subsection*{Technology model}
Since a regression model can only produce forecasts of data it can learn from, additional bottom-up processing must be carried out to get a full picture of India's demand in the future. We identify trends and data points at the state level of the country to build a regional profile as well as the national one. 

\subsubsection*{Cooling}
Cooling is divided into two main categories: residential and commercial. The ratio of commercial to residential consumption is computed from state-level data \cite{stats} and is used as the ratio of commercial to residential cooling demand. Using the IEA's baseline and efficient cooling projections from the Future of Cooling study \cite{iea}, we use the annual sales and unit types to calculate the energy consumption and growth rate at a national level and pro-rate it down to state level given GDP per capita. Surveyed hourly demand profiles \cite{usaid} are indicators of behavioral cooling energy consumption patterns as exemplified in Supplementary Fig. \ref{fig:sup-ac-res} and \ref{fig:sup-ac-com}. The survey produce various profiles given climate seasons, household income and size. We apply a time-domain convolution of these profiles to generate a representative profile for each state for the various climates and seasons.

We can generate the air conditioning demand profiles for two weather seasons (winter and summer) by convolution of the sample profiles to generate a smooth aggregated demand profile. Moreover, coincidence factors must be applied to properly estimate the simultaneity of the demand and its peak. Two coincidence factors are identified: weekday and weekend, values are extracted from a Reference Network Model Toolkit \cite{5504171}. We break down the national cooling demand to residential and commercial at state level by identifying state-level sector size and growth trends. Scaling the profiles to match the projected cooling energy demand produces hourly energy consumption profiles from residential and commercial cooling. Aggregating the appropriate states together will produce the same results at the regional level.

More importantly, the IEA’s future of cooling study \cite{iea} stresses the usage of Cooling Degree Days (CDD) to project cooling demand dependency on temperature. The unit consumption pattern and projections of capacity for India’s share of global cooling demand is based on growth in electrification, urbanization as well as Purchasing Power Parity. The IEA future of cooling study estimates that a 1-degree Celsius increase in decadal average temperature in 2050 will to lead to 25\% more CDD and a 2-degree Celsius increase will lead to 50\% more CDD. Climate change impacts are considered in the unit sales and energy consumption data used from the IEA’s future of cooling study. In our analysis, we use IEA's 50\% increase in CDD to model cooling demand in 2050. For prior periods, we interpolate CDD between 2018 and 2050 to model cooling demand. The increase in CDD and the addition of noise variation are introduced for the purpose of modeling the projected increase in peak demand due to climate change. Specifically, this analysis does not consider frequency nor forecast of extreme weather events.

\subsubsection*{Electric vehicles}
The second component of the technology model projects EV demand in India. The data presented here considered electric two, three, and four-wheel vehicles. Two-wheelers, being the dominating vehicle in terms of annual sales in India \cite{vehicle_sales}, are expected to be electrified first, followed by the three-wheelers and regular cars \cite{ey}. The Indian government has set a goal of converting 100\% of two-wheeler sales and 30\% of all vehicle sales to electric by 2030 \cite{nitiaayog}, so the starting point is vehicle sales at the state level \cite{vehicle_sales}. Using the regression equations of the corresponding GDP growth scenarios, we can project car sales with the EV targets by 2030 met in the rapid growth scenario. From vehicle sales and conversion rates, we get an estimate of the number of EV that will require charging. From a market survey on the average commute distance of vehicles in urban areas and rural areas \cite{ey}, long and short-range battery capacity and EV energy can be estimated. We introduce a mix of EV sales starting with short-range as the dominant market product and shifting to long-range, a market-dominant market in 2050. This trends reflects the current economic competitiveness of short-range EVs vs. existing internal combustion engine vehicles  as well as the long-term competitiveness of long-range EVs with declining battery costs.

Similar to the construction of the cooling profiles, a coincidence factor must be implemented, so as to not over-predict peak EV charging demand. Since this is a new consumption behavior and given the relatively small batteries of two-wheelers and three-wheelers, it is assumed that every vehicle needs to charge every other day on average for urban drivers and every day for rural ones. This yields an average daily consumption from EV charging. As shown in Supplementary Fig. \ref{fig:sup-ev-profiles}, three different charging profiles  --- home, work, public -- are identified in an EV pilot project study in Mexico City \cite{berkeley}. While Mexico and India differ greatly in many socio-economic aspects. The different hourly EV charging profiles collected were for a pilot project to deploy electric two-wheelers and small sedans in the metropolitan area of Mexico City. This presents two synergies enabling the usage of the charging profiles in India. Under the assumptions that EV deployment will be more prevalent in urban areas in India with initial conversion of smaller vehicles (two-wheelers and three-wheelers), the charging data collected \cite{berkeley} is a suitable fit for potential EV charging schemes in India. Energy consumption is computed from vehicle sales, projections, and electrification conversion. That calculated number is then fitted under the chosen charging profile. Time domain convolution of the profiles is applied to smoothen the peakiness of the total constructed hourly time series.

\subsubsection*{Data Dependence}

The technology model relies heavily on surveyed data to produce the representative hourly profiles for cooling and electric vehicle demands at state levels. This is indeed a limitation, and our projections assumes that future technology adopters will behave just like initial adopters. In the absence of a better alternative at a similar spatial and temporal resolution, the bottom-up modeling effort provides a reasonable estimate of temporal patterns expected from these new demand sources. For the hourly sample cooling profiles, the main assumption is that cooling demand consumption is only dependent on weather patterns and econometric patterns. Specifically, we apply a weighted sum convolution of the income level cooling profiles based on the states' GDP per capita ranking. For the total cooling demand at national level, we depend on the air cooling unit sales projection as well as break down of unit energy consumption under baseline and efficient scenarios of the IEA's Future of Cooling report \cite {iea}. We pro-rate residential cooling at state level using the GDP per capita projections. For commercial cooling we use the state-wise sector growth trends \cite{energystatistics}. A sanity check for this break down is to sum both residential and commercial state-wise cooling demand and compare to the IEA's all India cooling demand annual electricity consumption projections to 2050, the difference is highlighted in Supplementary Fig. \ref{fig:sup-ac_compare} and \ref{fig:sup-cooling}. Regarding the EV profiles, while there are alternative choices of charging schemes, we identified the synergies with the Berkeley study \cite{berkeley} to be best reflective of the bookend EV charging scenarios across India.

\section*{Data Records}
The data is uploaded on Zenodo \cite{marc_barbar_2020_4564581} and is available to download at \hyperlink{https://doi.org/10.5281/zenodo.4564581}{https://doi.org/10.5281/zenodo.4564581}. The path leading to a CSV file indicate the scenario corresponding to the results of that file. Breakdown of the folder hierarchy listed as:

\begin{enumerate}
    \item GDP Growth: slow, stable, rapid
    \item EV charging: home, work, public
    \item Cooling: baseline, efficient
    \item Type: detailed, summary
\end{enumerate}

The \textit{detailed} results are tables of the itemized hourly demand profile of each considered scenario; all files will produce 8760 rows (number of hours in a year). The \textit{summary} are tables of the itemized annual energy consumption for the considered years; all files will produce seven rows (number of considered future years). Both file types are itemized the same way as per Table \ref{table:headers}. The path of each file is the reference to the specific scenario the data in the tables represents. For example \textit{SR.csv} file under \textit{slow/home/efficient/summary} is the summary file of the case of slow economic growth, home EV and energy efficient air conditioning consumption.

\section*{Technical Validation}
The Business-as-usual statistical model is validated using standard statistical metrics when backtesting is applied. Further details on the backtesting are available elsewhere \cite{meia}. For the technology model, we compare our estimates to the IEA's WEO \cite{weo20,weo19,weo18,weo17} and Brookings India \cite{brookings}. Furthermore, our projections compare favorably against the EV projections to the IEA's Global Electric Vehicle Outlook 2020 \cite{ev}.

\subsection*{Back testing}

\begin{figure}[!ht]
\centering
 \includegraphics[width=7cm,height=7cm, keepaspectratio]{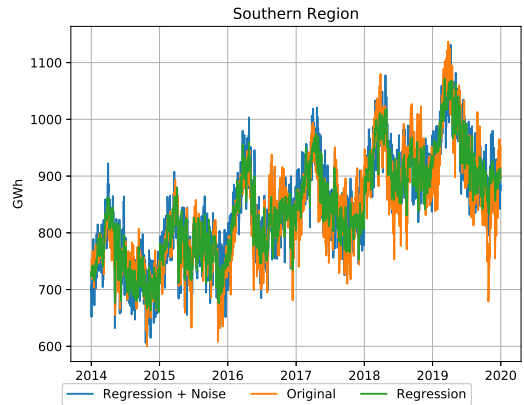}
  \caption{Back test result for Southern Region regression model}
   \label{fig:regression}
\end{figure}

Daily consumption and peak are projected for all five regions, we show the daily consumption back tests of the Southern Region in Fig. \ref{fig:regression}. More results can be found on the GitHub repository. It is important to note that the regression model captures the organic growth of the historical demand as well as the seasonal variation in demand but is not accurate at predicting daily variation. This shortcoming can be attributed to the small training dataset that is available. To compensate for this short-coming, we add additional noise variation as discussed earlier in the Methods section. We compare the R-squared value of the regression only versus the regression and noise time series as shown in Table \ref{table:r2}. Additionally, selected parameter performance metrics of the model for the Southern Region are presented in Table \ref{table:params}. The model's independent variables are the 2 meters and 10 meters elevation historic temperature and humidity data for the selected cities and GDP data for the state. Various weather parameters will have a higher coefficient then GDP since the latter is not as granular as a metric but will still be factored in for longer term growth as interpreted by its Fourier component \cite{meia}.

\subsection*{Cross-comparison}
Supplementary Fig. \ref{fig:sup-stated-weo} and \ref{fig:sup-sustainable-weo} compare the forecasting results to the WEO 2020 projections of India's Energy Demand to 2040. Our band of projections is notably wider due to the large number of scenarios that are combined to forecast energy demand. We further compare our results to Brookings India's study in Supplementary Fig. \ref{fig:sup-brookings}. We also compare our electric vehicle projections to those of the Global EV Outlook in Supplementary Fig. \ref{fig:sup-ev}. Finally, we compare our air conditioning demand contribution to the peak demand to the Future of Cooling study in Supplementary Fig. \ref{fig:sup-ac_compare}.

\subsection*{COVID-19 pandemic impact on year 2020}
\begin{figure}[!ht]
\centering
 \includegraphics[width=12cm,height=12cm, keepaspectratio]{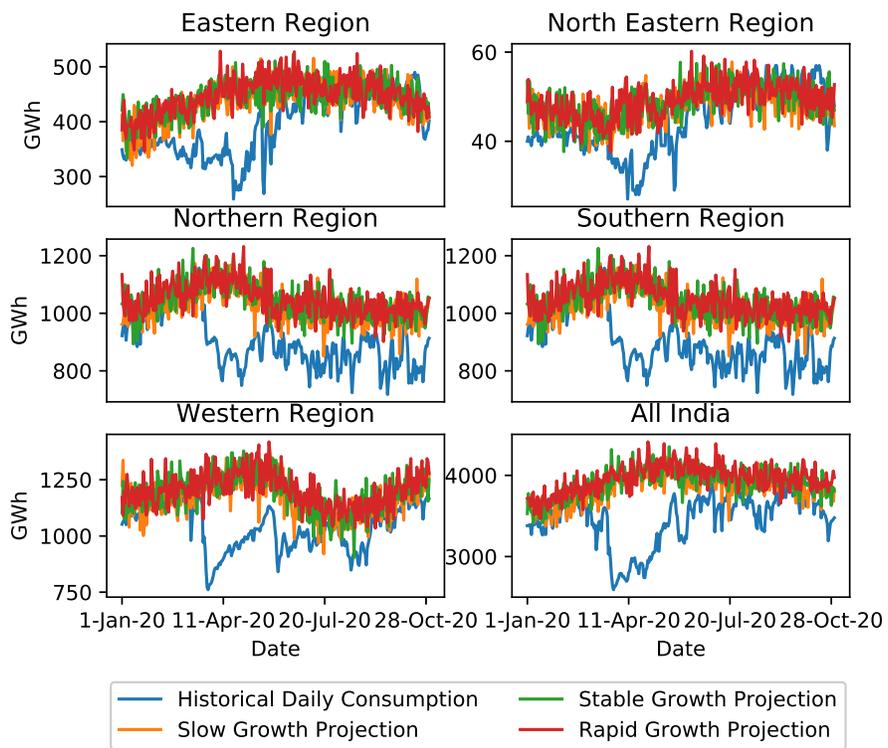}
  \caption{2020 year-to-date demand comparison with projections}
   \label{fig:comparison}
\end{figure}

The COVID-19 pandemic has drastically affected the global population in various ways. Energy consumption dropped severely as people were advised to stay at home. While it is not possible to project such "Black Swan" events from historical data, their long-term effects can be modeled as delayed growth under various recovery schemes. Fig. \ref{fig:comparison} shows that our projections for the month of January 2020 align with the realized demand, which is prior to the global outbreak of COVID-19. Evidently, there is a strong mismatch in the following months as the outbreak developed into a global pandemic. However, in the later part of the year, signs of recovery are noticed where the historical daily consumption once again reaches projected levels.

The impact of extreme events on energy consumption are difficult to predict at a granular level. Our projections are at a five year increment so that such yearly variations are smoothed out and the regression towards the mean phenomenon is observed. Moreover, the recovery from extreme events and their long-term impact can depend on many factors: economic, social, scientific and more. Without modeling those events in detail, projected growth can model the long-term average growth rate. In case of a negative extreme event, a smaller growth rate can model the long-term impact caused by the slow down. Similarly, a positive extreme event can be modeled as larger growth rate to include the long-term impact by the rapid growth. With signals of a fast recovery in total daily consumption for most regions, we elected to disregard projections that model long-term COVID-19 pandemic impact to avoid confirmation bias. Moreover, there is little data to support projections modeling a long-term impact on Indian energy consumption. We believe that the model and data presented in this paper are valid beyond the COVID-19 pandemic.

\section*{Usage Notes}

The format of the results is comma-separated values (CSV). All the results are available on the Zenodo Open-Access repository \cite{marc_barbar_2020_4564581}.

\section*{Code availability}

The code used in the generation of the data sets is open-sourced on Github repository \cite{git}.

\bibliography{sample}

\begin{thebibliography}{10}
\urlstyle{rm}
\expandafter\ifx\csname url\endcsname\relax
  \def\url#1{\texttt{#1}}\fi
\expandafter\ifx\csname urlprefix\endcsname\relax\def\urlprefix{URL }\fi
\expandafter\ifx\csname doiprefix\endcsname\relax\def\doiprefix{DOI: }\fi
\providecommand{\bibinfo}[2]{#2}
\providecommand{\eprint}[2][]{\url{#2}}

\bibitem{weo20}
\bibinfo{author}{{International Energy Agency}}.
\newblock \emph{\bibinfo{title}{World Energy Outlook 2020}}
  (\bibinfo{publisher}{IEA}, \bibinfo{year}{2020}).

\bibitem{ev}
\bibinfo{author}{{International Energy Agency}}.
\newblock \emph{\bibinfo{title}{Global EV Outlook 2020}}
  (\bibinfo{publisher}{IEA}, \bibinfo{year}{2020}).

\bibitem{iea}
\bibinfo{author}{{International Energy Agency}}.
\newblock \emph{\bibinfo{title}{The Future of Cooling}}
  (\bibinfo{publisher}{IEA}, \bibinfo{year}{2018}).

\bibitem{teri}
\bibinfo{author}{Spencer, T.} \& \bibinfo{author}{Awasthy, A.}
\newblock \emph{\bibinfo{title}{\emph{Analysing and projecting Indian
  electricity demand to 2030}}} (\bibinfo{publisher}{The Energy and Resources
  Institute}, \bibinfo{year}{2019}).

\bibitem{brookings}
\bibinfo{author}{Ali, S.}
\newblock \emph{\bibinfo{title}{\emph{The future of Indian electricity demand:
  how much, by whom and under what conditions?}}} (\bibinfo{publisher}{The
  Brookings Institute}, \bibinfo{year}{2018}).

\bibitem{mospi}
\bibinfo{author}{{Ministry of Statistics and Programme Implementation}}.
\newblock \bibinfo{title}{\emph{National accounts statistics 2019}}.
\newblock \bibinfo{howpublished}{\url{http://mospi.nic.in/}}
  (\bibinfo{year}{2019}).

\bibitem{shakti}
\bibinfo{author}{{Shakti Sustainable Energy Foundation, GE India Exports Pvt.
  Ltd}}.
\newblock \bibinfo{title}{\emph{Modelling with power sector planning for
  India}} (\bibinfo{year}{2018}).

\bibitem{ivan}
\bibinfo{author}{García, I.~R.}
\newblock \emph{\bibinfo{title}{Decarbonizing the Indian power sector by 2037 :
  evaluating different pathways that meet long-term emissions targets}}.
\newblock Master's thesis, \bibinfo{school}{Massachusetts Institute of
  Technology} (\bibinfo{year}{2019}).

\bibitem{mckinsey}
\bibinfo{author}{Sankhe, S.} \emph{et~al.}
\newblock \bibinfo{title}{\emph{India’s turning point: an economic agenda to
  spur growth and jobs}} (\bibinfo{year}{2020}).

\bibitem{pwc}
\bibinfo{author}{PricewaterhouseCoopers}.
\newblock \bibinfo{title}{\emph{The long view how will the global economic
  order change by 2050?}} (\bibinfo{year}{2017}).

\bibitem{gompertz}
\bibinfo{author}{Winsor, C.~P.}
\newblock \bibinfo{journal}{\bibinfo{title}{The gompertz curve as a growth
  curve}}.
\newblock {\emph{\JournalTitle{Proceedings of the National Academy of Sciences
  of the United States of America}}} \textbf{\bibinfo{volume}{18}},
  \bibinfo{pages}{1--8} (\bibinfo{year}{1932}).

\bibitem{ssrn}
\bibinfo{author}{Datta, P.}
\newblock \bibinfo{title}{{Population projection of EAG states of India: vision
  until 2051 Preprint at \url{https://doi.org/10.2139/ssrn.186176}}}
  (\bibinfo{year}{2011}).
\newblock \bibinfo{note}{Submitted}.

\bibitem{eia}
\bibinfo{author}{{United States Energy Information Administration}}.
\newblock \bibinfo{title}{\emph{India’s economic growth is driving its energy
  consumption}} (\bibinfo{year}{2013}).

\bibitem{posoco}
\bibinfo{author}{{Ministry of Power}}.
\newblock \bibinfo{title}{\emph{Power system operation corporation: load
  dispatch center monthly reports}}.
\newblock \bibinfo{howpublished}{\url{https://posoco.in/reports/}}
  (\bibinfo{year}{2020}).

\bibitem{nasa}
\bibinfo{author}{Gelaro, R.} \emph{et~al.}
\newblock \bibinfo{journal}{\bibinfo{title}{The modern-era retrospective
  analysis for research and applications, version 2 ({MERRA}-2)}}.
\newblock {\emph{\JournalTitle{Journal of Climate}}}
  \textbf{\bibinfo{volume}{30}}, \bibinfo{pages}{5419--5454}
  (\bibinfo{year}{2017}).

\bibitem{meia}
\bibinfo{author}{Alsup, M.}
\newblock \emph{\bibinfo{title}{Forecasting electricity demand in the data-poor
  Indian context}}.
\newblock Master's thesis, \bibinfo{school}{Massachusetts Institute of
  Technology} (\bibinfo{year}{2020}).

\bibitem{pop}
\bibinfo{author}{{United Nations, Department of Economic and Social Affairs,
  Population Division}}.
\newblock \emph{\bibinfo{title}{World Population Prospects 2019: Highlights}}.
\newblock \bibinfo{number}{{Report No. ST/ESA/SER.A/423}}
  (\bibinfo{publisher}{{United Nations}}, \bibinfo{year}{2019}).

\bibitem{scikit-learn}
\bibinfo{author}{Pedregosa, F.} \emph{et~al.}
\newblock \bibinfo{journal}{\bibinfo{title}{Scikit-learn: Machine learning in
  {P}ython}}.
\newblock {\emph{\JournalTitle{Journal of Machine Learning Research}}}
  \textbf{\bibinfo{volume}{12}}, \bibinfo{pages}{2825--2830}
  (\bibinfo{year}{2011}).

\bibitem{git}
\bibinfo{author}{Barbar, M.}
\newblock \bibinfo{title}{Source code for: Demand forecasting india}.
\newblock \bibinfo{howpublished}{\emph{Github}.
  \url{https://github.com/barbarmarc/demand-forecasting-india/}}
  (\bibinfo{year}{2020}).

\bibitem{usaid}
\bibinfo{author}{Garg, A.}, \bibinfo{author}{Maheshwari, J.} \&
  \bibinfo{author}{Upadhyay, J.}
\newblock \emph{\bibinfo{title}{Load Research for Residential and Commercial
  Establishments in Gujarat}}, vol. \bibinfo{volume}{ECO-III-1024} of
  \emph{\bibinfo{series}{Energy Conservation and Commercialization}}
  (\bibinfo{publisher}{Indian Institute of Management Ahmedabad},
  \bibinfo{year}{2010}).

\bibitem{weo19}
\bibinfo{author}{{International Energy Agency}}.
\newblock \emph{\bibinfo{title}{World Energy Outlook 2019}}
  (\bibinfo{publisher}{IEA}, \bibinfo{year}{2019}).

\bibitem{stats}
\bibinfo{author}{{Datanet India Development Team}}.
\newblock \bibinfo{title}{\emph{Electrical energy consumption by industry}}.
\newblock \bibinfo{howpublished}{\url{https://www.indiastat.com/}}
  (\bibinfo{year}{2020}).

\bibitem{5504171}
\bibinfo{author}{{Mateo Domingo}, C.}, \bibinfo{author}{{Gomez San Roman}, T.},
  \bibinfo{author}{{Sanchez-Miralles}, A.}, \bibinfo{author}{{Peco Gonzalez},
  J.~P.} \& \bibinfo{author}{{Candela Martinez}, A.}
\newblock \bibinfo{journal}{\bibinfo{title}{A reference network model for
  large-scale distribution planning with automatic street map generation}}.
\newblock {\emph{\JournalTitle{IEEE Transactions on Power Systems}}}
  \textbf{\bibinfo{volume}{26}}, \bibinfo{pages}{190--197}
  (\bibinfo{year}{2011}).

\bibitem{vehicle_sales}
\bibinfo{author}{{Society of India Automobile Manufacturers}}.
\newblock \bibinfo{title}{\emph{Automobile domestic sales trends 2015-2020}}.
\newblock \bibinfo{howpublished}{\url{https://www.siam.in/}}
  (\bibinfo{year}{2020}).

\bibitem{ey}
\bibinfo{author}{Batra, R.} \emph{et~al.}
\newblock \emph{\bibinfo{title}{Standing up India’s EV ecosystem - who will
  drive the charge?}} (\bibinfo{publisher}{Ernest and Young},
  \bibinfo{year}{2018}).

\bibitem{nitiaayog}
\bibinfo{author}{{Government of India National Institution for Transforming
  India}}.
\newblock \bibinfo{title}{\emph{Zero emission vehicles: towards a policy
  framework}}.
\newblock
  \bibinfo{howpublished}{\url{https://niti.gov.in/writereaddata/files/document_publication/EV_report.pdf}}
  (\bibinfo{year}{2018}).

\bibitem{berkeley}
\bibinfo{author}{von Meier, A.}
\newblock \bibinfo{title}{{Ev infrastructure planning and grid impact
  assessment: a case for Mexico Preprint at
  \url{http://www2.eecs.berkeley.edu/Pubs/TechRpts/2018/EECS-2018-70.html}}}
  (\bibinfo{year}{2018}).
\newblock \bibinfo{note}{Submitted}.

\bibitem{energystatistics}
\bibinfo{author}{{Ministry of Statistics and Programme Implementation}}.
\newblock \bibinfo{title}{\emph{Energy statistics 2019}}.
\newblock \bibinfo{howpublished}{\url{http://mospi.nic.in/}}
  (\bibinfo{year}{2019}).

\bibitem{marc_barbar_2020_4564581}
\bibinfo{author}{Barbar, M.}, \bibinfo{author}{Mallapragada, D.},
  \bibinfo{author}{Alsup, M.} \& \bibinfo{author}{Stoner, R.}
\newblock \bibinfo{title}{Electricity demand forecasting of india}.
\newblock \bibinfo{howpublished}{\emph{Zenodo}.
  \url{https://doi.org/10.5281/zenodo.4564581}} (\bibinfo{year}{2020}).

\bibitem{weo18}
\bibinfo{author}{{International Energy Agency}}.
\newblock \emph{\bibinfo{title}{World Energy Outlook 2018}}
  (\bibinfo{publisher}{IEA}, \bibinfo{year}{2018}).

\bibitem{weo17}
\bibinfo{author}{{International Energy Agency}}.
\newblock \emph{\bibinfo{title}{World Energy Outlook 2017}}
  (\bibinfo{publisher}{IEA}, \bibinfo{year}{2017}).

\end{thebibliography}

\section*{Acknowledgements}

This research was supported by MIT Energy Initiative's Low Carbon Energy Centers and the Future of Storage Study

\section*{Author contributions statement}

R.S. collected the data, M.B. and M.A. built the Technology Model, M.B. built the Business-as-usual Model, and D.M. validated the results. All authors reviewed the manuscript. 

\section*{Competing interests} 

The authors declare no competing interests.

\section*{Figures \& Tables}

\begin{table}[!ht]
\centering
\begin{tabular}{|c | c|} 
 \hline
 Data & Source \\ [0.5ex] 
 \hline\hline
 State-wise Historical GDP & Ministry of Statistics and Programme Implementation \cite{mospi}\\ 
 Vehicle Sales and Registration & Society of indian Automobile Manufacturers \cite{vehicle_sales} \\
 Air Conditioning Stock and Capacity & International Energy Agency \cite{iea} \\
 Load Profile & United States Agency for International Development \cite{usaid} \\
 State-level sector-wise energy consumption & Power System Operation Corporation \cite{posoco,energystatistics} \\ [1ex]
 \hline
\end{tabular}
\caption{Input Data Sources}
\label{table:data}
\end{table}

\begin{table}[!ht]
\centering
\begin{tabular}{|c | c|} 
 \hline
 Slow & Rapid \\ [0.5ex] 
 \hline\hline
Gompertz Growth Curve & Exponential Growth Curve \\
\begin{math}y = A e^{-e^{\frac{\mu e (B - x)}{A}} + 1} + C\end{math}& \begin{math}y = A e^{Bx} +C\end{math}\\
$$A = \num{7e-29}$$ & $$A = \num{1e-64}$$\\
$$B = \num{4.64e-2}$$ & $$B = \num{8.7e-2}$$\\
$$C=0$$&$$C=0$$\\
R-squared = 0.949669 & R-squared = 0.989361\\
 \hline
\end{tabular}
\caption{GDP projections curve fit results}
\label{table:gdp}
\end{table}

\begin{table}[!ht]
\centering
\begin{tabular}{|c | c|} 
 \hline
 Column Header & Description \\ 
 \hline\hline
 DateTime & Hourly or yearly time resolution\\
 Base & Business-as-usual model resulting demand\\
 Com AC & Commercial Air Conditioning demand\\
 Res AC & Residential Air Conditioning demand\\
 E2W & Electric Two-Wheelers demand\\
 E3W & Electric Three-Wheelers demand\\
 E4W & Electric Four-Wheelers demand\\
 \hline
\end{tabular}
\caption{Output data headers descriptor}
\label{table:headers}
\end{table}

\begin{table}[!ht]
\centering
\begin{tabular}{|c | c c|} 
 \hline
 Region & Regression & Regression + Noise \\ 
 \hline\hline
 Eastern & 0.709 & 0.798\\
 Northeastern & 0.608 & 0.722\\
 Northern & 0.691 & 0.784\\
 Southern & 0.744 & 0.825\\
 Western & 0.680 & 0.778\\
 \hline
\end{tabular}
\caption{Business-as-usual Regression R-squared consumption results}
\label{table:r2}
\end{table}

\begin{table}[!ht]
\centering
\begin{tabular}{|c | c c c c|} 
 \hline
  Parameter&Coefficient&Standard Error&t value&P > | t |\\ 
 \hline\hline
GDP&13.3630&34.663&0.386&0.700\\
Bangalore t2m max       &  931.0014 &   353.052   &   2.637   &   0.008     \\
Chennai h2m min &         1.736e+04 &  1.65e+04   &   1.050   &   0.294\\
Hyderabad t10m max     &  -502.7828  &  356.558 &    -1.410 &     0.159\\
Vijayawada h10m min   &   -2.229e+04 &  7671.803   &  -2.905   &   0.004\\
 \hline
\end{tabular}
\caption{Business-as-usual Southern Region consumption Regression performance of select parameters}
\label{table:params}
\end{table}

\end{document}


\listoffigures

\vspace{20cm}

\begin{figure}[!ht]
\centering
 \includegraphics[width=9cm,height=9cm, keepaspectratio]{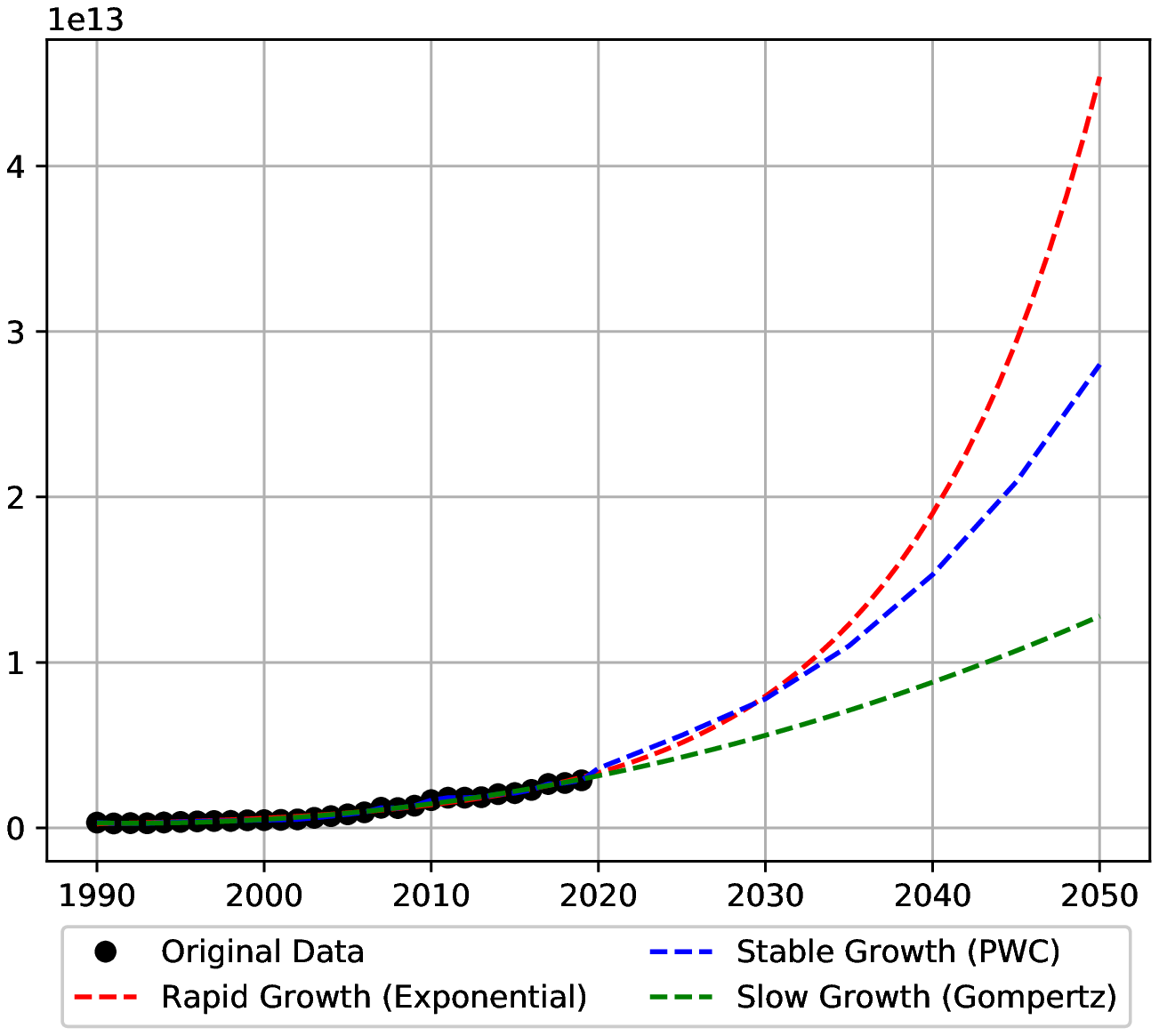}
  \caption{India's GDP curve-fit and forecasting to 2050}
   \label{fig:sup-gdp}
\end{figure}

\begin{figure}[!ht]
\centering
 \includegraphics[width=9cm,height=9cm, keepaspectratio]{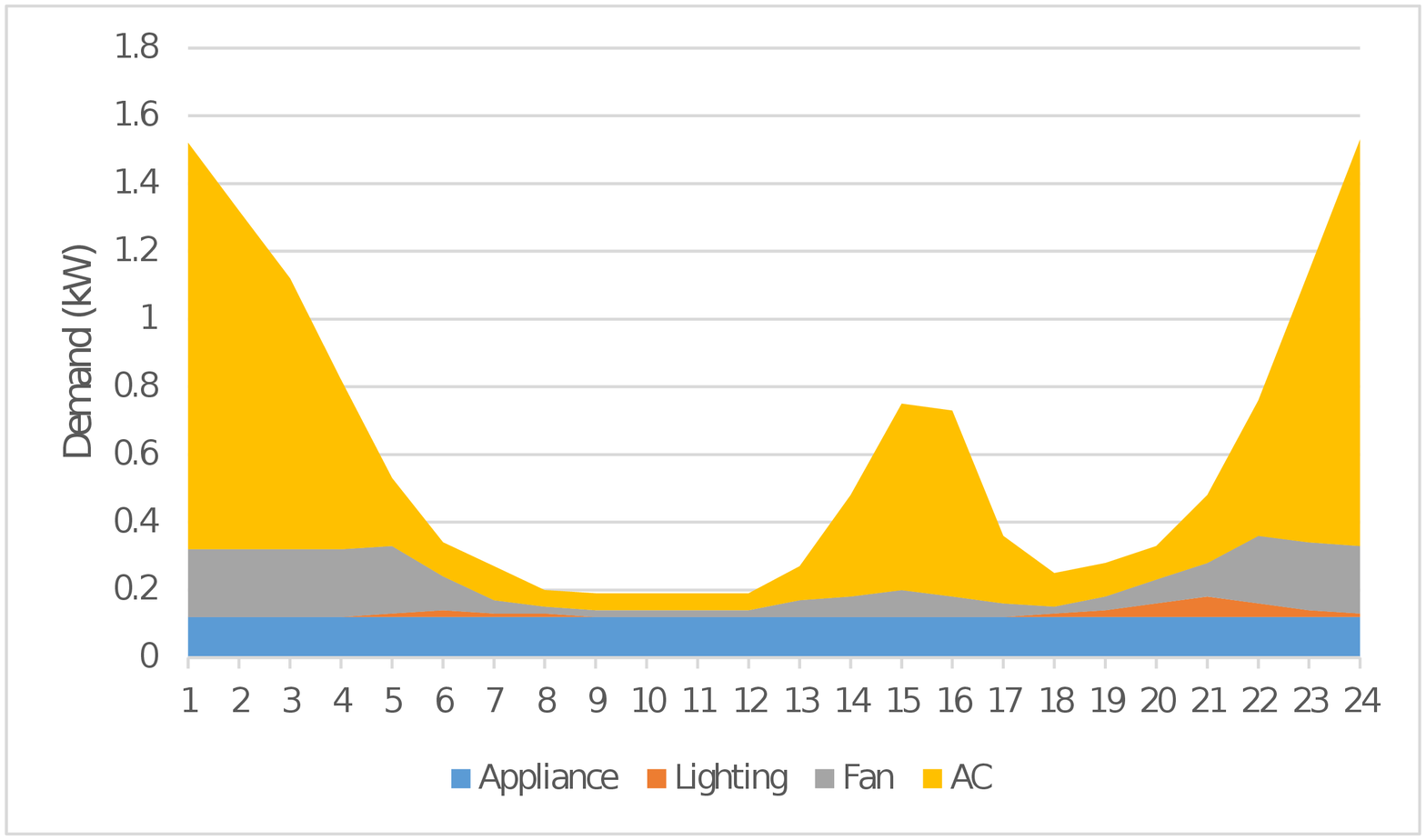}
  \caption{Residential survey categorized hourly demand profile}
   \label{fig:sup-ac-res}
\end{figure}

\begin{figure}[!ht]
\centering
 \includegraphics[width=9cm,height=9cm, keepaspectratio]{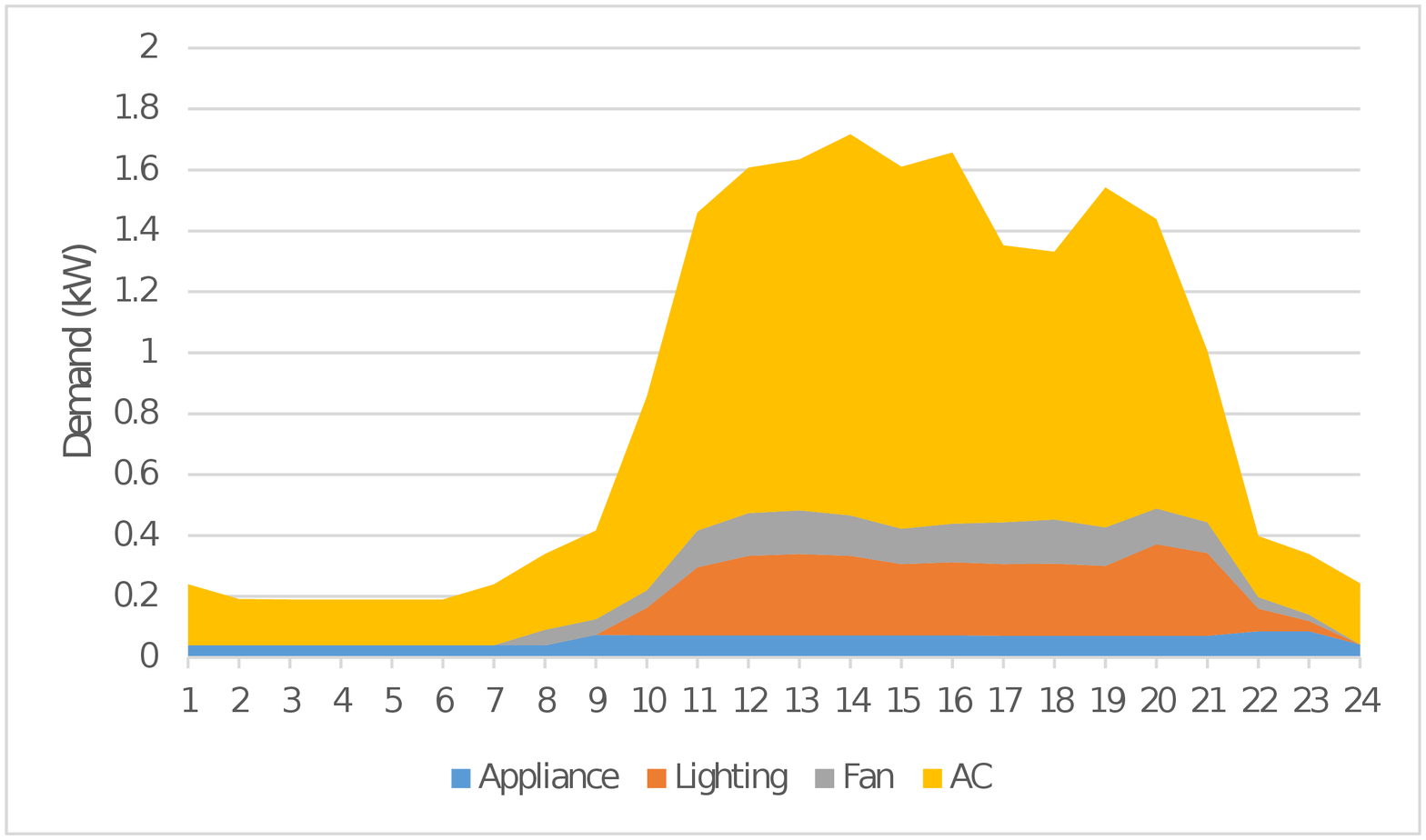}
  \caption{Commercial survey categorized hourly demand profile}
   \label{fig:sup-ac-com}
\end{figure}

\begin{figure}[!ht]
\centering
 \includegraphics[width=9cm,height=9cm, keepaspectratio]{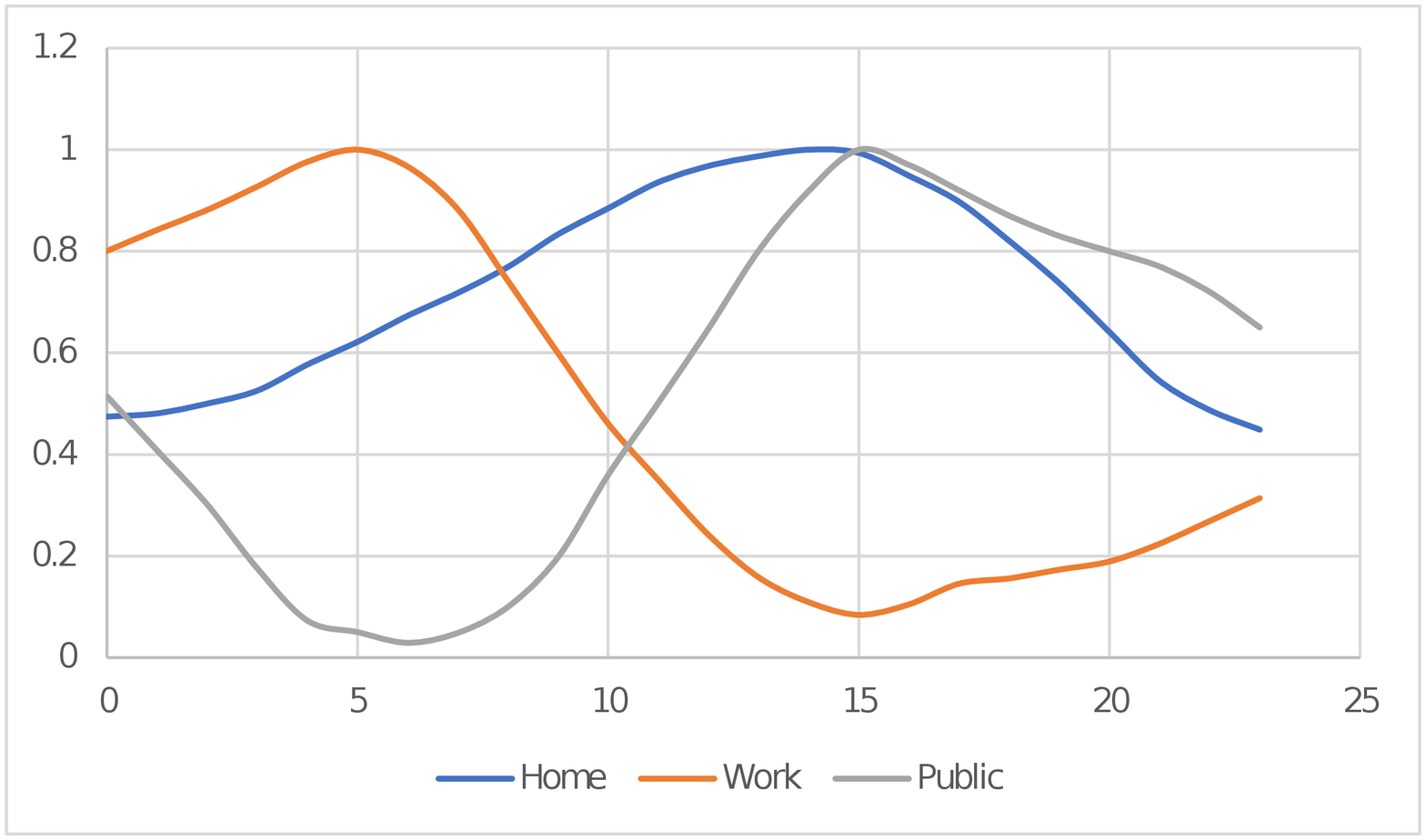}
  \caption{Normalized sample charging profile schemes}
   \label{fig:sup-ev-profiles}
\end{figure}

\begin{figure}[!ht]
\centering
 \includegraphics[width=9cm,height=9cm, keepaspectratio]{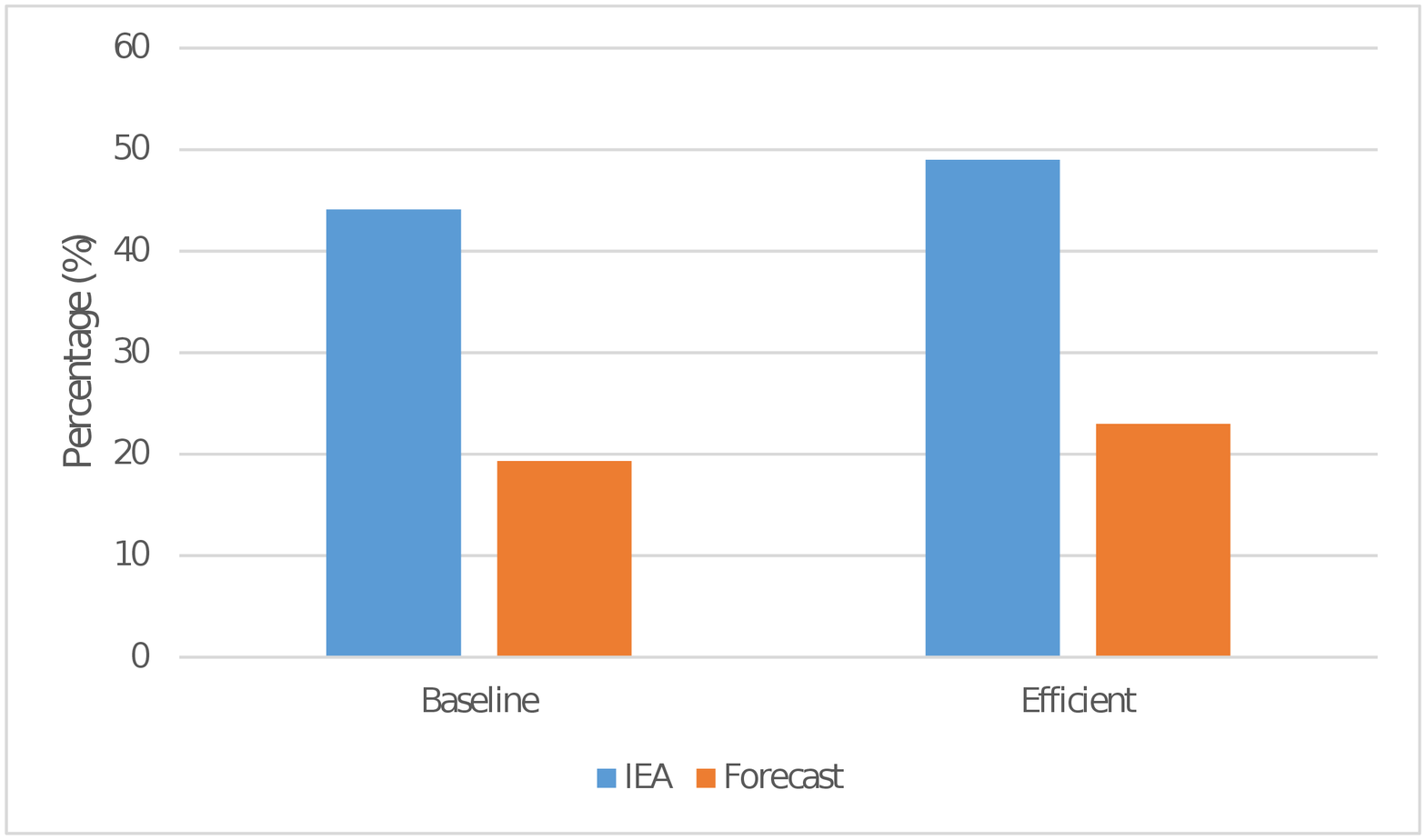}
\caption{2050 cooling demand contribution to peak results comparison with IEA's Future of Cooling}
\label{fig:sup-ac_compare}
\end{figure}

\begin{figure}[!ht]
\centering
 \includegraphics[width=9cm,height=9cm, keepaspectratio]{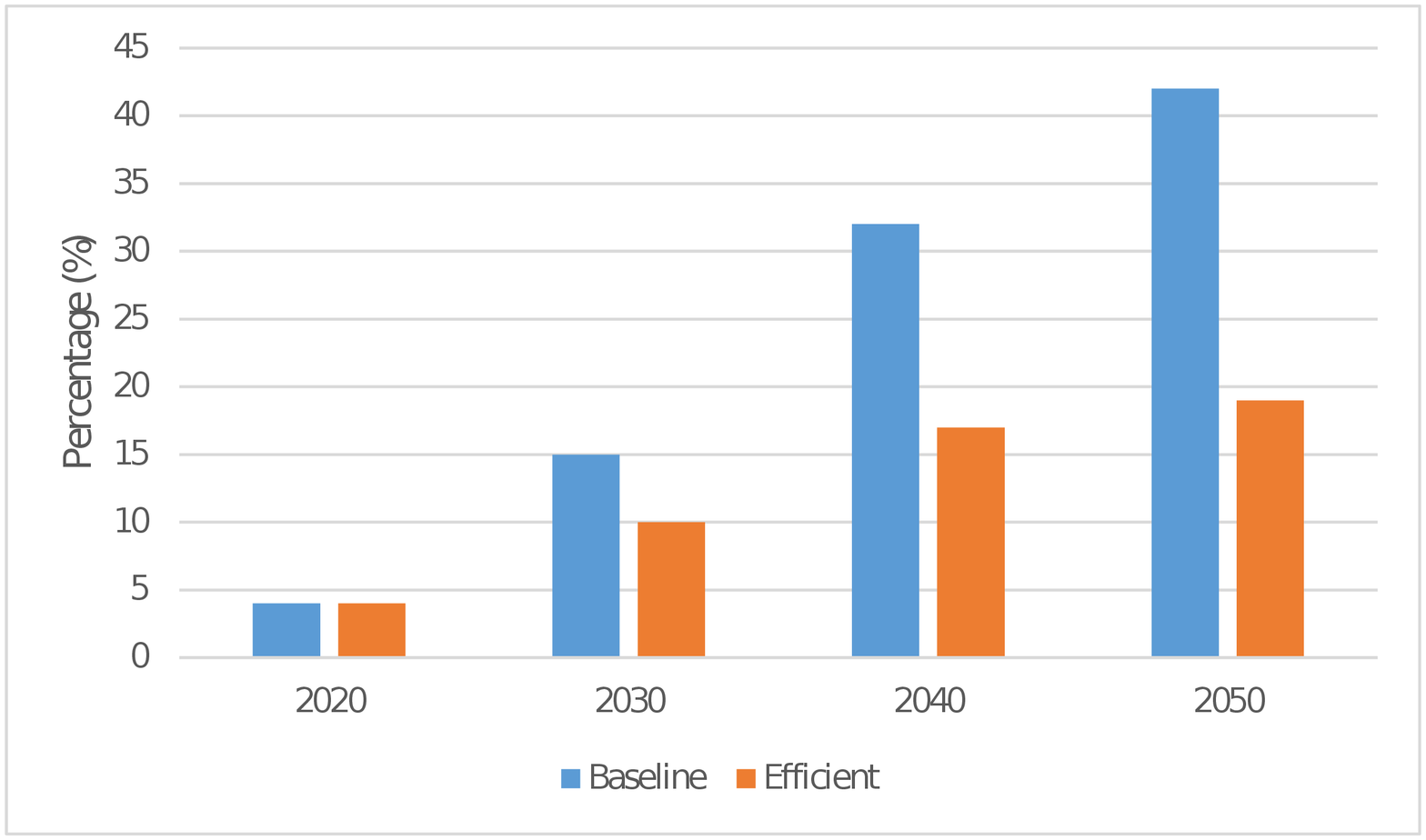}
\caption{Cooling demand contribution to peak demand}
\label{fig:sup-cooling}
\end{figure}

\begin{figure}[!ht]
\centering
 \includegraphics[width=9cm,height=9cm, keepaspectratio]{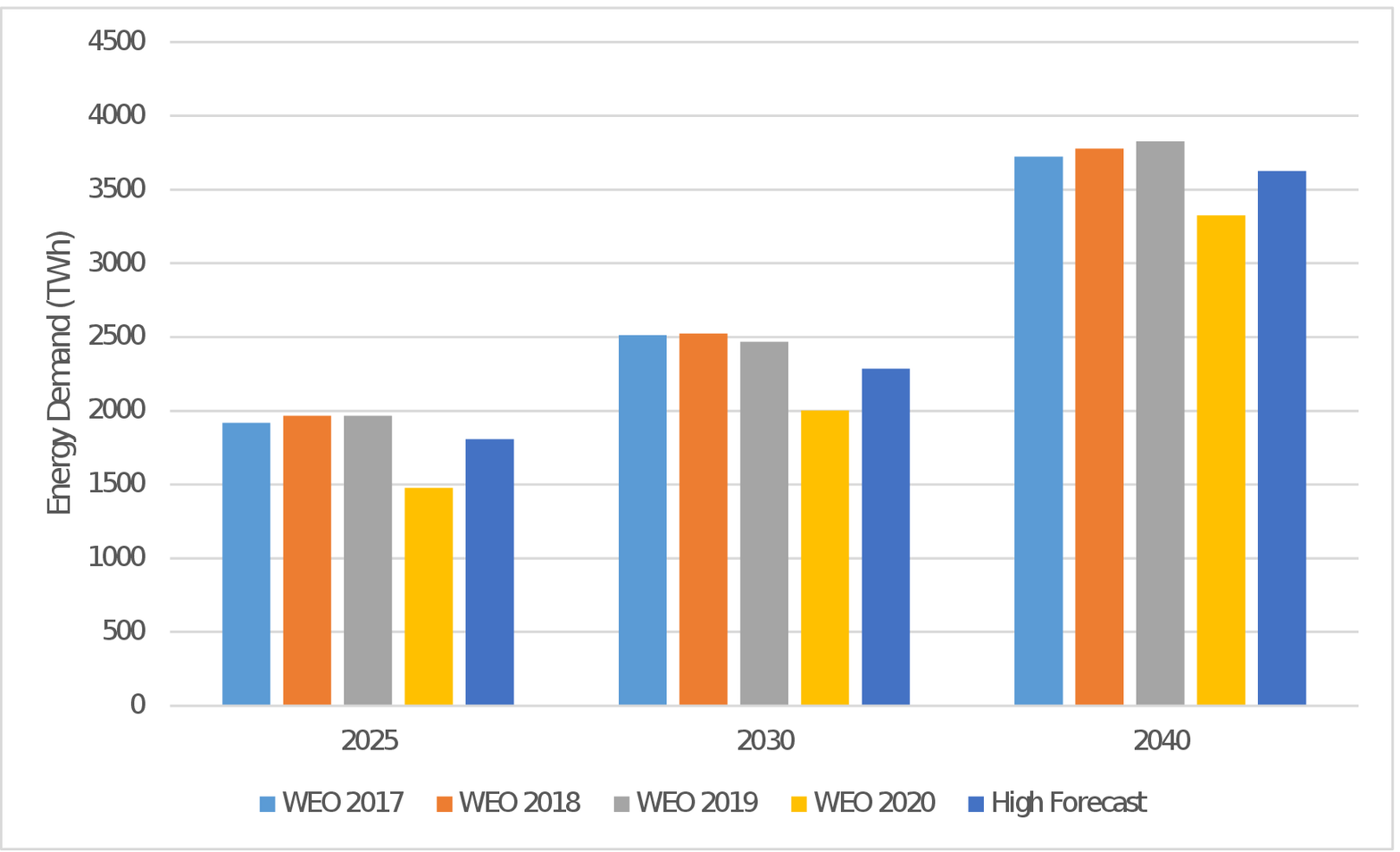}
\caption{Results comparison with stated policy World Energy Outlook projections}
\label{fig:sup-stated-weo}
\end{figure}

\begin{figure}[!ht]
\centering
 \includegraphics[width=9cm,height=9cm, keepaspectratio]{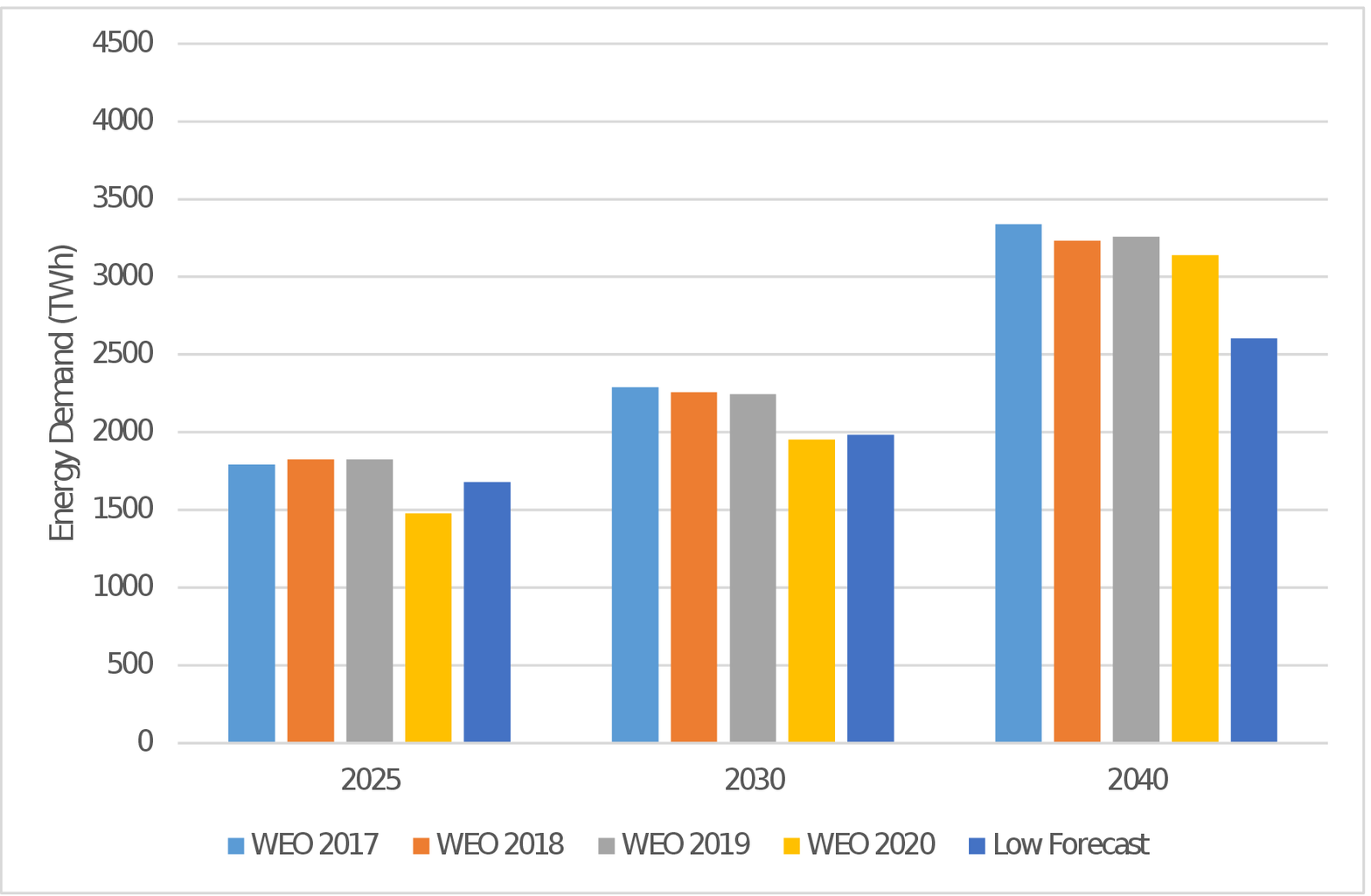}
\caption{Results comparison with sustainable policy World Energy Outlook projections}
\label{fig:sup-sustainable-weo}
\end{figure}

\begin{figure}[!ht]
\centering
 \includegraphics[width=9cm,height=9cm, keepaspectratio]{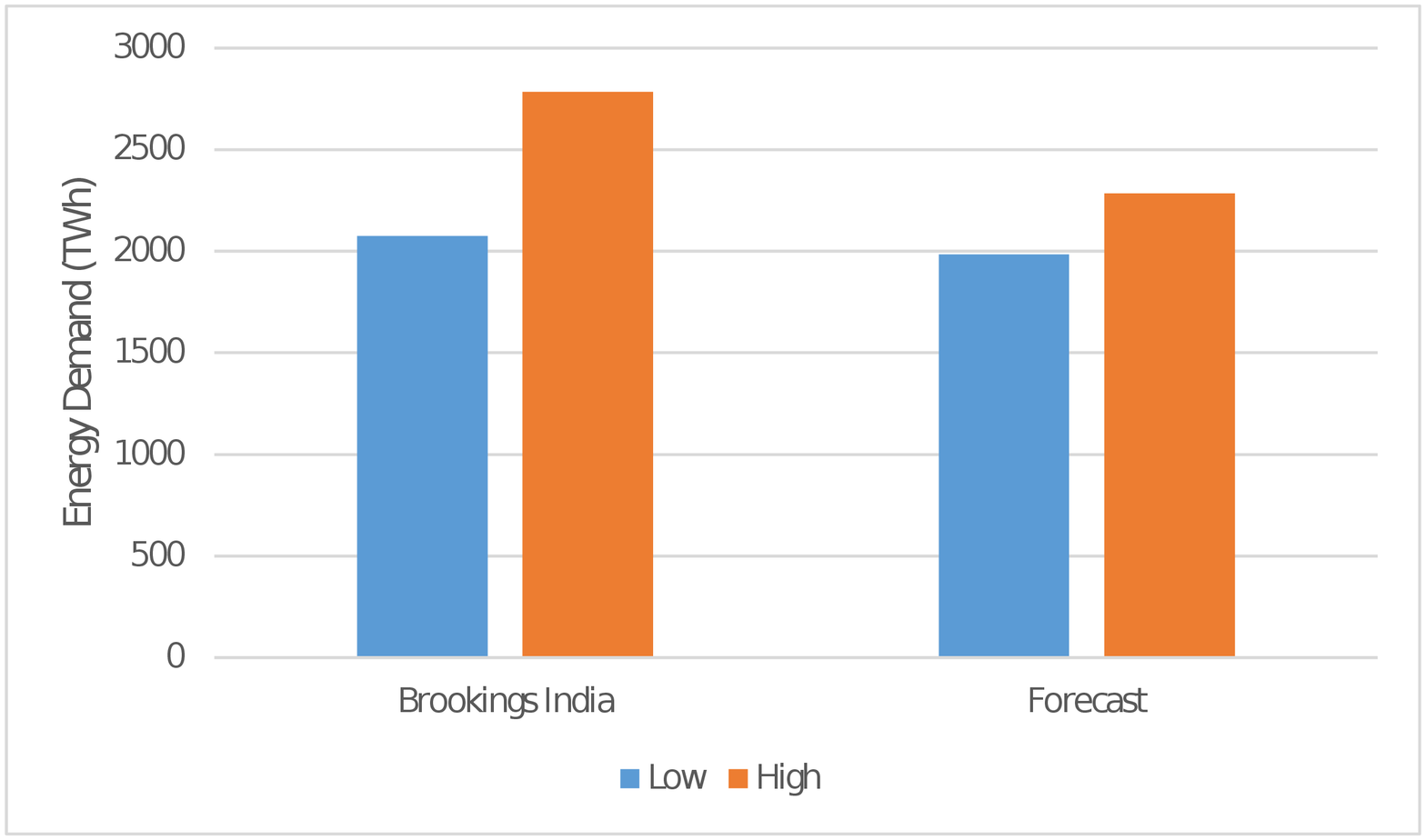}
\caption{Results comparison with Brookings India 2030 projections}
\label{fig:sup-brookings}
\end{figure}

\begin{figure}[!ht]
\centering
 \includegraphics[width=9cm,height=9cm, keepaspectratio]{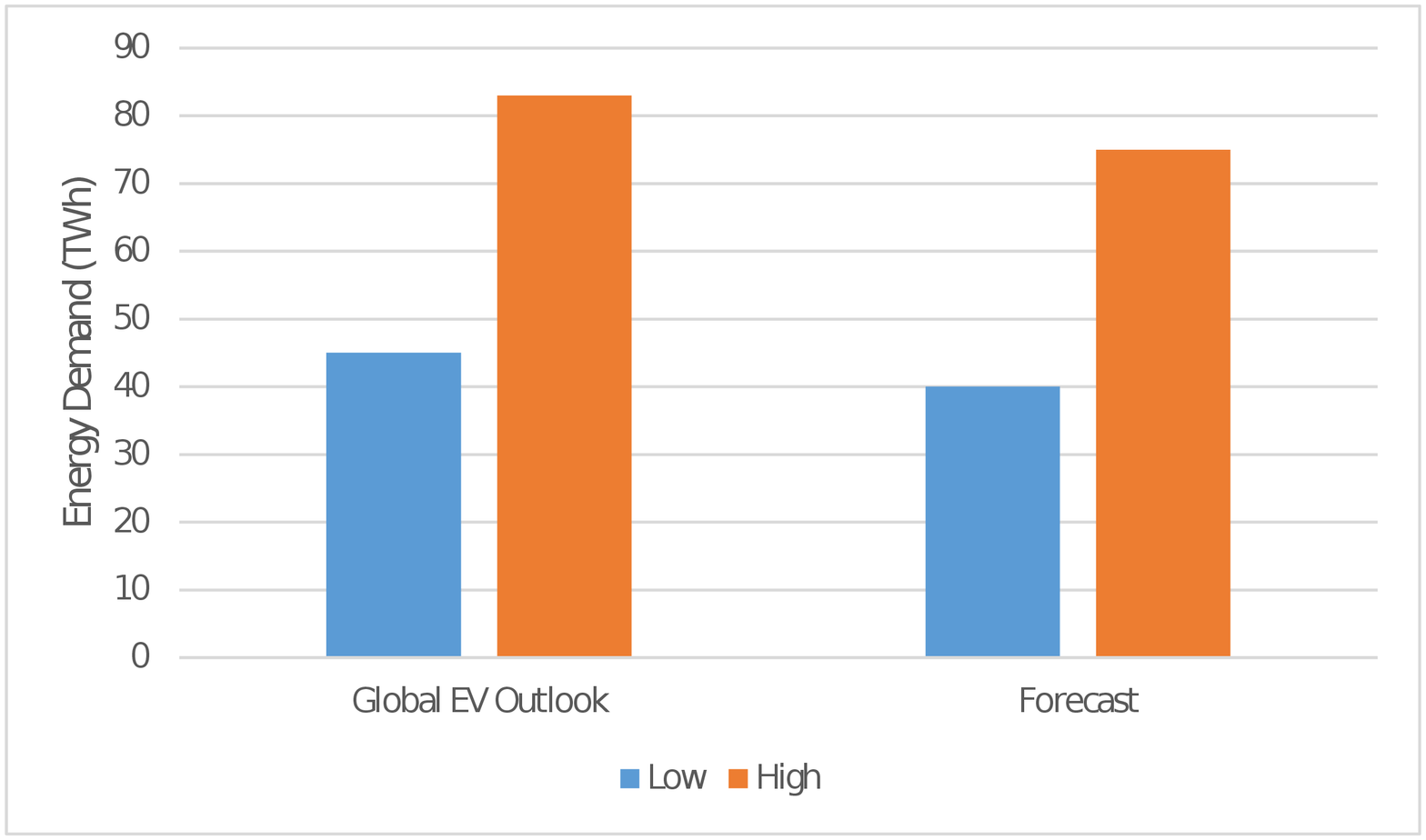}
\caption{Electric Vehicle demand results comparison with IEA's Global EV Outloook}
\label{fig:sup-ev}
\end{figure}